\documentclass[twocolumn,prl,aps,showpacs,preprintnumbers,superscriptaddress]{revtex4}
\usepackage{amsthm,amsfonts,graphicx,verbatim}
\newcommand{\be}{\begin{equation}}
\newcommand{\ee}{\end{equation}}
\newcommand{\bea}{\begin{eqnarray}}
\newcommand{\eea}{\end{eqnarray}}

\newcommand{\lp}{\left(}
\newcommand{\rp}{\right)}

\renewcommand{\epsilon}{\varepsilon}
\renewcommand{\vec}[1]{{\bf #1}}

\begin{document}

\title{Long-Range Interaction Between Adatoms in Graphene}

\author{Andrei V. Shytov}
\affiliation{Department of Physics,
University of Utah,
Salt Lake City, UT 84112}
\author{Dmitry A. Abanin}
\affiliation{Department of Physics, Jadwin Hall, Princeton University, Princeton, NJ 08544}
\author{Leonid S. Levitov}
\affiliation{Department of Physics, Massachusetts Institute of Technology, Cambridge MA 02139}
\begin{abstract}
We present a theory of electron-mediated interaction between adatoms in graphene. In the case of resonant scattering, relevant for hydrogentated graphene, a long-range $1/r$ interaction is found. This interaction can be viewed as a fermionic analog of the Casimir interaction, in which massless fermions play the role of photons. The interaction is an attraction or a repulsion depending on whether the adatoms reside on the same sublattice or on different sublattices, with attraction dominating for adatoms randomly distributed over both sublattices. The attractive nature of these forces creates an instability under which adatoms tend to aggregate. 
\end{abstract}
\maketitle

Unique transport characteristics of graphene make it a strong candidate for replacing silicon in future electronic devices \cite{Geim07}. 
Functionalizing graphene by controllable oxidation\,\cite{Stankovich06,Liu08} or hydrogenation\,\cite{Ryu08,Elias08} can change its properties in new, unexpected ways. In particular, when hydrogen adatoms bind to graphene, the orbital state of each functionalized carbon atom changes from $sp^2$ to $sp^3$ configuration, removing $\pi$-electrons from the conduction band and turning graphene into a semiconductor \cite{Duplock04}. Remarkably, the metallic properties can be fully recovered after dehydrogenation \cite{Elias08}. 
This provides a unique tool to control electronic properties of this material
\cite{Sofo07,Boukhvalov08}. 


One of interesting questions posed by the experiment \cite{Elias08} has to do with the interaction between adatoms mediated by electron scattering. As we shall see, resonant scattering on the midgap states localized on adatoms \cite{Pepin01,Bena05,Pereira06} leads to dramatic enhancement of interaction, making it long-ranged.
We find that the interaction energy falls off very slowly, approximately as inverse distance between the adatoms, $U(r)\sim r^{-1}$. The sign of interaction depends on the sublattice type: two atoms residing on different sublattices ($A$ and $B$) attract, whereas atoms on the same sublattice repel [see Eqs.(\ref{eq:interaction_AB}),(\ref{eq:AA_final})]. 

The $r^{-1}$ interaction is stronger than the long-range interaction between adatoms on surfaces of metals \cite{Crommie93,Avouris94,Repp00}, which is of a Friedel-oscillation (FO) character. The FO interaction falls off as $r^{-2}$ when it is mediated by electronic states on the surface, and as $r^{-3}$, when mediated by the states in the bulk \cite{Einstein73,Lau78}. The FO interaction can occur in graphene \cite{Cheianov06}. Long-range interaction can lead to fascinating collective behavior of adatoms, such as
self-organization into chains \cite{Repp00} and superlattices \cite{Silly04}.

The interaction analyzed in this work can be interpreted as a fermionic Casimir effect. 
The Casimir interaction between two bodies (or, atoms) 
arises due to scattering of virtual photons.
For each of the bodies, in the presence of the second body, angular distribution of the flux of incident virtual photons is somewhat anisotropic, giving rise to a net attraction force. 
This interaction is of a generic character (fermionic Casimir effect was recently analyzed in one-dimensional systems \cite{Zhabinskaya08}).

\begin{figure}
\includegraphics*[width=3.4in]{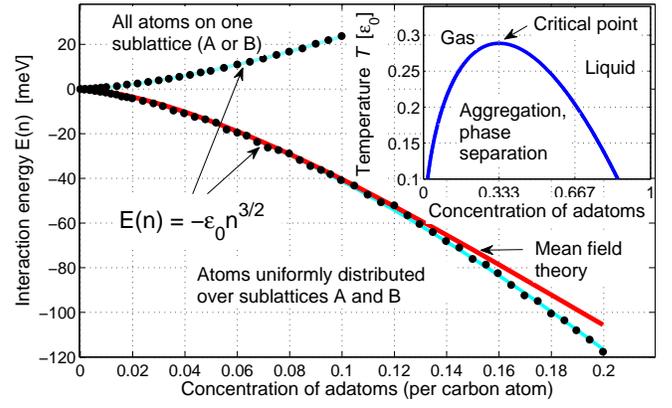}
\caption{
Electron-mediated interaction between adatoms in graphene modeled by a hard-core potential:
numerical results (black dots) and mean field theory, Eq.(\ref{eq:Omega_int}) (red line). The net interaction is a repulsion when adatoms are randomly placed on one of the sublattices, and an attraction when they are equally distributed over both sublattices. 
The $3/2$ power law (\ref{eq:E(n)}) provides an accurate fit to the numerical results with the best fit values $\epsilon_0=-0.75\,{\rm eV}$ (top curve) and $\epsilon_0=1.3\,{\rm eV}$ (bottom curve).
System of size $48\times 82$ was used for simulation, each data point was averaged over 20 realizations of randomly generated adatom configurations. 
Inset: Attracting adatoms tend to aggregate.
Phase diagram obtained from the free energy (\ref{eq:free_energy}) is shown.
}
\label{fig2}
\end{figure}

We find that attraction between atoms on different sublattices is stronger by a logarithmic factor than repulsion within the same sublattice. The net interaction of atoms equally distributed among the two sublattices 
is thus an attraction, characterized by the energy density
\be\label{eq:E(n)}
E(n)= -\epsilon_0 n^{3/2}
,\quad \epsilon_0\approx 1.3\,{\rm eV},
\ee
per carbon atom (see Fig.\ref{fig2}), where $n$ is the fraction of hydrogenated carbon atoms.
The prefactor in (\ref{eq:E(n)}) may have a weak logarithmic dependence on $n$.

We emphasize that the interaction energy in this case cannot be treated as a sum of pairwise two-particle interactions (indeed, summing $1/r$ interactions over the entire space would give a divergence). The situation resembles that of Casimir forces, which are of an essentially non-pairwise nature. To treat
the interaction mediated by electrons one must 
account for the change in electronic states at the energies $\epsilon\lesssim \hbar v_0 n^{1/2}$, resulting from electron scattering on the adatoms ($v_0$ is the electron Fermi velocity). This leads to interaction energy per adatom of order $\hbar v_0/r$, with 
$r=n^{-1/2}$ the typical distance between adatoms, in agreement with $n^{3/2}$ scaling, Eq.(\ref{eq:E(n)}).

Attraction can lead to instability of a homogeneous phase and 
adatom aggregation.
Characteristic time scales for such processes are controled by the rates of adsorption and desorption, or diffusion, whichever is faster.
Compression of the graphene lattice, resulting from attraction between adatoms, may expalain the observed
reduction of the lattice constant\,\cite{Elias08}, 
which is at odds with first-principles calculations\,\cite{Boukhvalov08}.

Interaction between hydrogen adatoms could also result from corrugation of the graphene sheet caused by the stress around tetrahedral $sp^3$ bonds. Numerical evidence suggests, however, that such corrugation is limited to the range of at most a few lattice constants\cite{Boukhvalov08}, rendering this type of interaction effectively short-ranged.

\begin{figure}
\includegraphics*[width=1.6in]{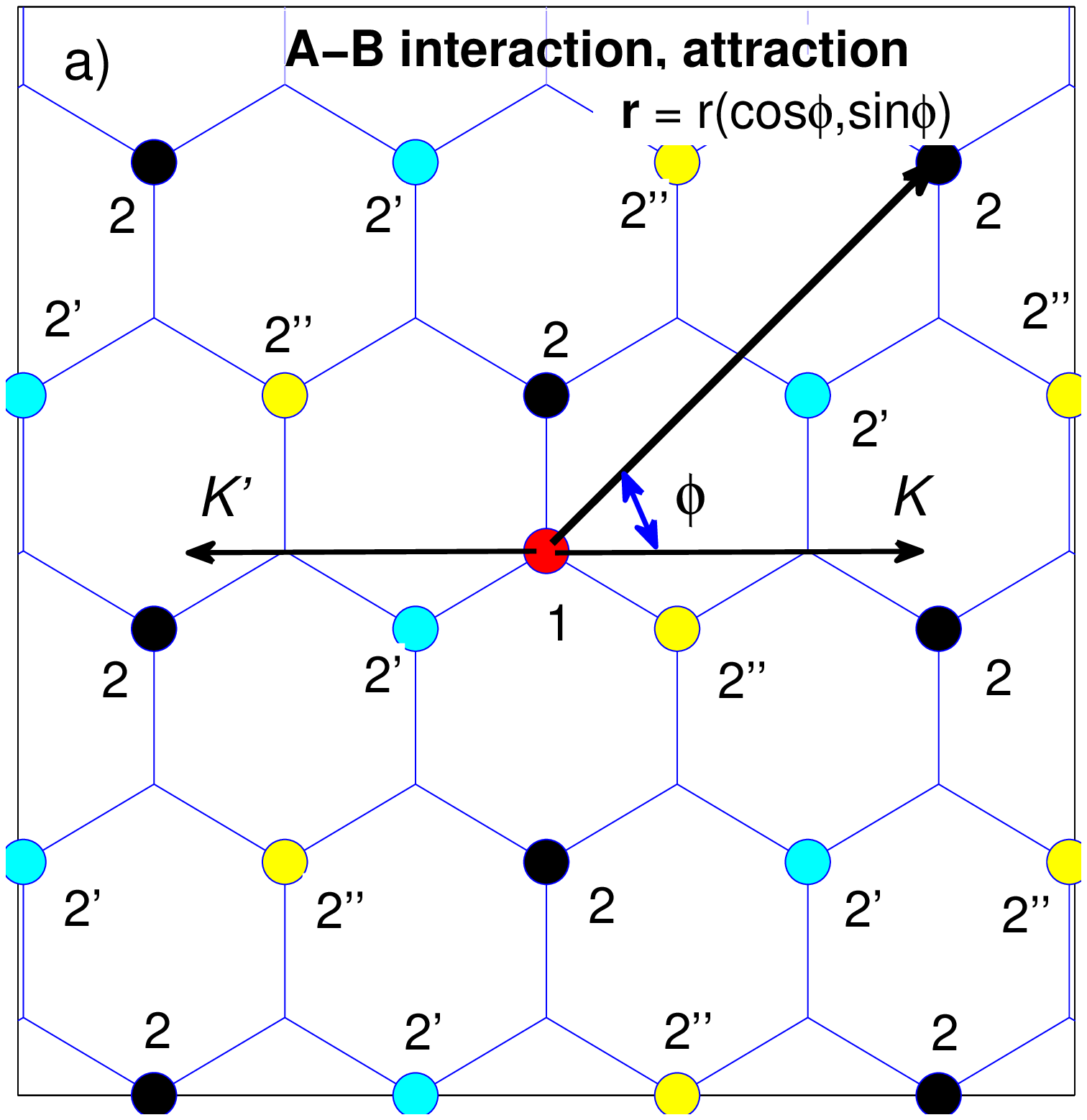}
\includegraphics*[width=1.6in]{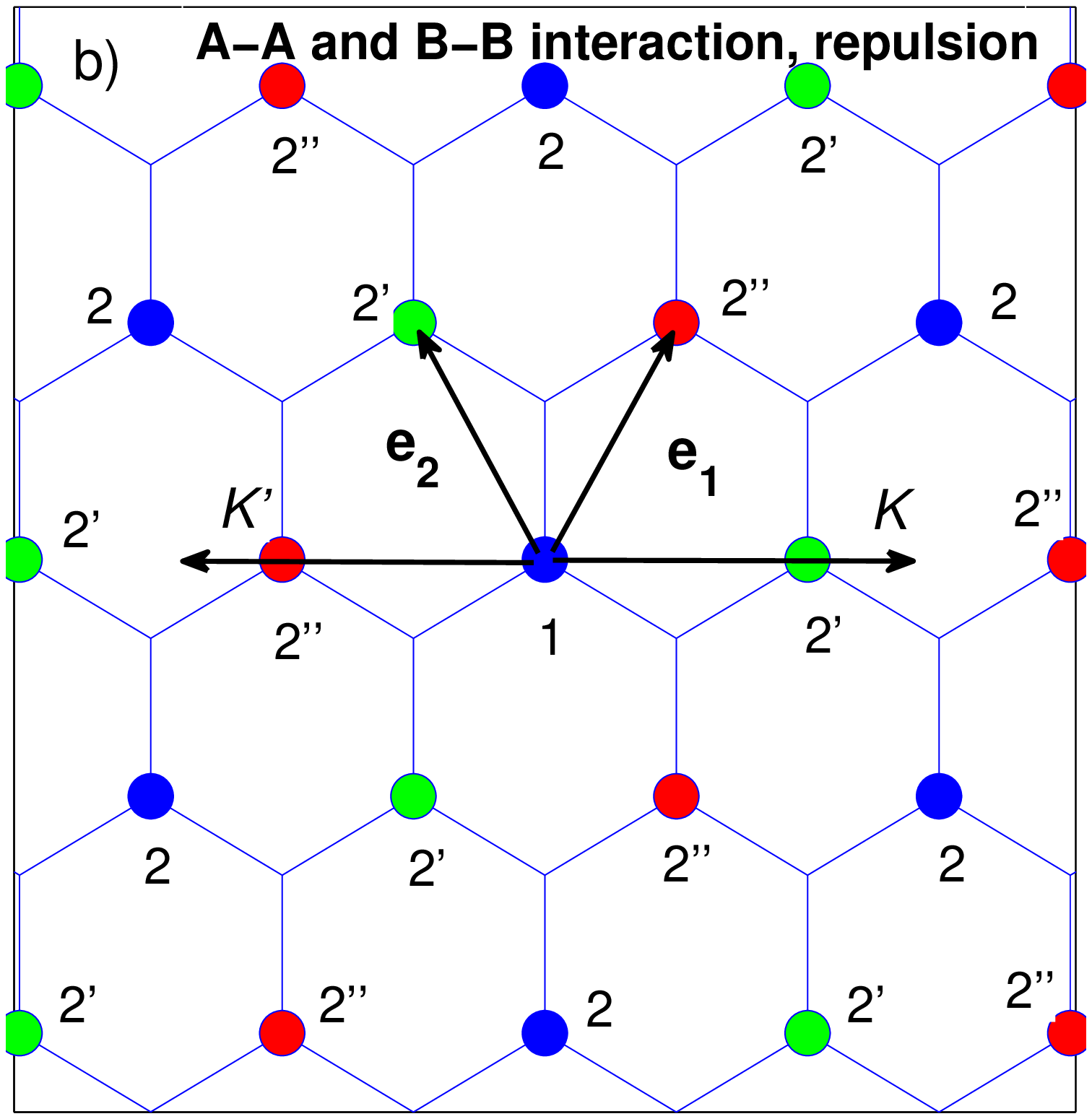}
\caption{Electron-mediated interaction between adatoms depends on the type of sublattice: atoms on different sublattices, $A$ and $B$, attract (a), whereas atoms on the same sublattice repel (b)
[see Eq.(\ref{eq:interaction_AB}) and Eq.(\ref{eq:AA_final})]. The interaction 
is modulated by a prefactor 
which takes different values on the three sub-sublattices marked by $2$, $2'$ and $2''$: (a) $|\sin(\vec K\vec r+\phi)|=|\sin\phi|,\,|\sin(\phi+\frac{2\pi}3)|,\,|\sin(\phi-\frac{2\pi}3)|$; (b) $\cos^2(\vec K\vec r)=1,\,\frac14,\,\frac14$. The modulation results from interference between electronic states in valleys $K$ and $K'$.
}
\label{fig1}
\end{figure}

The problem of electrons scattering on impurities can be described by a tight-binding Hamiltonian
%
\be\label{eq:H}
H=\sum_{\vec k}( t_{\vec k}\psi^\dagger_{\vec k,A}\psi_{\vec k,B}+{\rm h.c.}) + \!\!\! \sum_{\vec x,\,\alpha=A,B}\!\!\! u_\alpha(\vec x)\psi^\dagger_{\vec x,\alpha}\psi_{\vec x,\alpha}
.
\ee
Here $u_{A(B)}(\vec x)$ is adatoms' potential on sublattices $A$($B$), and $t_{\bf k}=t_0(1+e^{-i{\bf ke_1}}+e^{-i{\bf ke_2}})$, with $t_0\approx 3.1\,{\rm eV}$ the hopping amplitude and ${\bf e}_{1(2)}$ the basis vectors (see Fig.\ref{fig1}b).

The interaction between adatoms can be conveniently analyzed in the Matsubara Greens function framework using the thermodynamical potential $\Omega=T\sum_{\epsilon_n}{\rm Tr}\,\ln G$ \cite{AGD}. 
For two adatoms, we write $G^{-1}=G^{-1}_0-V_1(\vec x-\vec x_1)-V_2(\vec x-\vec x_2)$. Resumming perturbation series in terms of the T-matrices of each adatom, we obtain
%
\be\label{eq:Omega_12}
\Omega=-T\sum_{\epsilon_n}{\rm Tr}\,\ln \lp 1-T_1G_{12}T_2G_{21}\rp
.
\ee
Here $G_{12}$ is the free-particle Greens function in position representation, evaluated between the points $\vec x_1$ and $\vec x_2$ (similar representation was used recently in a study of Casimir forces \cite{Emig07,Kenneth08}).

In this paper we shall use the s-wave resonant scattering approximation,
\be\label{eq:Tmatrix}
T_0(i\epsilon)=
\frac{\pi v_0^2}{i\epsilon\ln(W/|\epsilon|)+\delta}
,\quad 
|\delta|\ll W\approx 3t_0
,
\ee
as appropriate for short-range scatterers at low energies. Here $W$ is the electron half-bandwidth, and the parameter $\delta$ describes detuning of resonance from the Dirac point. An expression of this form can be obtained for a delta-function potential,
$u(\vec x)=V\delta(\vec x-\vec x_1)$, in which case the T-matrix is given by $T(\epsilon)=V/\lp 1+\frac{V}{\pi v_0^2}i\epsilon\ln\frac{W}{|\epsilon|}\rp$ \cite{Pepin01,Bena05,Pereira06}. 
For hydrogen adsorbed on graphene,
the presence of a resonance peak close to the Dirac point,
Eq.(\ref{eq:Tmatrix}), was confirmed by first-principles calculations \cite{Duplock04}.

The real-space Greens function can be written as 
\be\label{eq:greens}
G(i\epsilon,\vec r)=-\int \frac{d^2k}{(2\pi)^2}\frac{e^{i\vec k\vec r}}{\epsilon^2+|t_{\vec k}|^2} \left[\begin{array}{cc}
         i\epsilon & t_{\vec k}  \\
         t^*_{\vec k} &  i\epsilon
      \end{array}\right],
\ee
where the matrix accounts for the $A$ and $B$ sublattices.
The Greens function takes on different form for the end points on different sublattices:
\be\label{eq:greens_low}
G(i\epsilon,\vec r)=
\left[\begin{array}{cc}
         G_{AA} & G_{AB}  \\
         G_{BA} &  G_{BB} \end{array} \right], 
\ee
In the low-energy approximation we expand $t_{\vec k}$ in the vicinity of points $\vec K$, $\vec K'=-\vec K$ to obtain
$t_{\vec k}\approx v_0(\mp p_x-ip_y)$, where $\vec p=\vec k\mp\vec K$ is the momentum relative to the $\vec K$ ($\vec K'$) point, and $v_0=\frac32t_0$ is the Fermi velocity.
Adding contributions of the states near $\vec K$ and $\vec K'$, we obtain 
\bea\label{eq:GAA}
&& G_{AA}=G_{BB}=-\frac{i\epsilon \cos(\vec K\vec r)}{\pi v_0^2} K_0\left(\epsilon \tilde r \right)
, \quad
\tilde r=\frac{r}{v_0}
,
\\\label{eq:GAB}
&& G_{AB}
=-\frac{\epsilon\sin ({\vec K\vec r+\phi)}}{\pi v_0^2} K_1\left(\epsilon \tilde r \right),
\eea
where $\phi$ is the angle between $\vec r$ and $\vec K$ (see Fig.\ref{fig1}a),
and $K_{0,1}$ denote modified Bessel functions of the second kind,
$K_\nu(z)=\frac{\Gamma(\nu+\frac12)2^\nu}{\sqrt{\pi}z^\nu}\int_0^\infty\frac{\cos zt\, dt}{(1+t^2)^{\nu+1/2}}$.
The function $G_{BA}$ can be obatined from the relation $G_{BA}(\vec r)=G_{AB}^*(-\vec r)$, giving
%
\be\label{eq:GBA}
G_{BA}=-\frac{\epsilon\sin ({\vec K\vec r-\phi)}}{\pi v_0^2} K_1\left(\epsilon \tilde r \right).
\ee
%
%
We first consider two adatoms on different sublattices
(see Fig.\ref{fig1}a). 
At distances $r\lesssim \hbar v_0/T$, 
approximating the sum in  $\Omega_{12}=-T\sum_{\epsilon_n}
\ln\lp 1-T^2_0(i\epsilon)G^2_{AB}(i\epsilon_n,\vec r)\rp $
by an integral $\int \frac{d\epsilon}{2\pi}$,
and using Eq.(\ref{eq:GAB}), 
we find
\be\label{eq:Omega_AB}
\Omega_{12}=-\int \frac{d\epsilon}{2\pi} \log\left(1-\frac{\epsilon^2\sin^2({\bf Kr}+\phi) K_1^2(\epsilon \tilde r)}{(i\epsilon \log(W /\epsilon)+\delta)^2}\right).
\ee
This result further simplifies for relatively short distances $r\lesssim\hbar v_0/\delta$.
The integral can be evaluated 
using the asymptotic formula $K_1(x\ll1)\approx 1/x$ and replacing 
$\ln(W/\epsilon)$ by $\ln(rW/\hbar v_0)$
with logarithmic accuracy. Setting $\delta=0$ and using the identity $\int_0^\infty dx\ln(1+u/x^2)=\pi\sqrt{u}$ we integrate over $\epsilon$ to obtain
\be\label{eq:interaction_AB}
U_{AB}(\tilde a\lesssim r\lesssim \hbar v_0/\delta)=\Omega_{12}\approx -\frac{\hbar v_0 |\sin(\vec K\vec r+\phi)|}{r \log (r/\tilde a)}
,
\ee
where $\tilde a=\hbar v_0/W$.
The interaction has a negative sign, corresponding to {\it attraction} of adatoms.

Interestingly, due to the factor $|\sin({\bf Kr}+\phi)|$  in the above expression, the interaction {\it oscillates on the lattice scale}. This oscillation results from interference of the contributions due to fermions from $K$ and $K'$ valleys.

The meaning of the factor $|\sin({\bf Kr}+\phi)|$ can be seen more clearly by considering it separately on each of the three sub-sublattices, which have period $\sqrt 3$ times the period of the $A$ or $B$ sublattice (see Fig.\ref{fig1}a). 
Since $e^{i{\bf Kr}}$ takes values $1$, $e^{2\pi i/3}$, and $e^{4\pi i/3}$, the same on each of the three sub-sublattices, the angular dependence in Eq.(\ref{eq:interaction_AB}) is given by $|\sin({\phi})|$, $|\sin(\phi+2\pi/3)|$, or $|\sin(\phi+4\pi/3)|$ in each of the three cases.

For a pair of adatoms residing on the same sublattice ($A$ or $B$),
the interaction energy is $\Omega_{12}=-T\sum_{\epsilon_n}\ln\lp 1-(T_0(i\epsilon_n)G_{AA}(i\epsilon_n,\vec r))^2\rp$, giving 
\be\label{eq:Omega_AA_integral}
\Omega_{12}=
-\int\frac{d\epsilon}{2\pi}\ln\lp 1+\frac{\epsilon^2\cos^2({\bf Kr}) K_0^2(\epsilon \tilde r)}{(i\epsilon\ln(W /\epsilon)+\delta)^2} \rp
.
\ee
We note a different sign under the log in this expression as compared to 
Eq.(\ref{eq:Omega_AB}), which arises because $G_{AA}$ is imaginary-valued, whereas $G_{AB}$ is real-valued.
The integral over $\epsilon$ is dominated by the region $\delta\lesssim |\epsilon|\lesssim v_0/r$, since $K_0(x)$ decreases exponentially for $x\gtrsim 1$.
For such $\epsilon$, and for $\ln(W  r/v_0)\gg 1$, the ratio $K_0(\epsilon \tilde r)/\ln(W /\epsilon)$ is small in most of the integration domain [$K_0(x\ll1)\approx -\log x$].
Thus we can Taylor-expand the log and, with logarithmic accuracy, integrate over $\epsilon$ using the identity $\int_0^\infty K_0^2(x)dx=\pi^2/4$, to obtain
\be\label{eq:AA_final}
U_{AA}(\tilde a\ll r\lesssim \hbar v_0/\delta )
\approx \frac{\pi \hbar v_0}{4r\log^2(r/\tilde a)}\cos^2({\bf Kr})
.
\ee
%
The factor $\cos^2 ({\bf Kr})$ in Eq.(\ref{eq:AA_final}), describing interference between two valleys, takes constant value on each of the three sub-sublattices with period $\sqrt 3$  (see Fig.\ref{fig1}b). 
Analyzing it as above we find that $\cos^2 ({\bf Kr})=1$
for adatoms residing on the same sub-sublattice, and $\cos^2 ({\bf Kr})=1/4$ when adatoms reside on different sub-sublattices.

The energy of interaction for adatoms on the same sublattice, Eq.(\ref{eq:AA_final}), is positive, which means that in this case adatoms {\it repel} each other. This repulsion is logarithmically weaker than the attraction found for atoms on different sublattices, Eq.(\ref{eq:interaction_AB}). We thus expect the net interaction for a system of many adatoms randomly placed on both sublattices to be dominated by attraction.

The repulsion (\ref{eq:AA_final}) will be greatest for the next-nearest carbon atoms. Interestingly, in an STM experiment \cite{Hornekaer06} it was found that chemisorbed hydrogen atoms can reside on the nearest or next-next-nearest sites of the carbon lattice, but never on the next-nearest sites.
This behavior is consistent with our results, Eqs.(\ref{eq:AA_final}),(\ref{eq:interaction_AB}). 


\begin{figure}
\includegraphics*[width=3.4in]{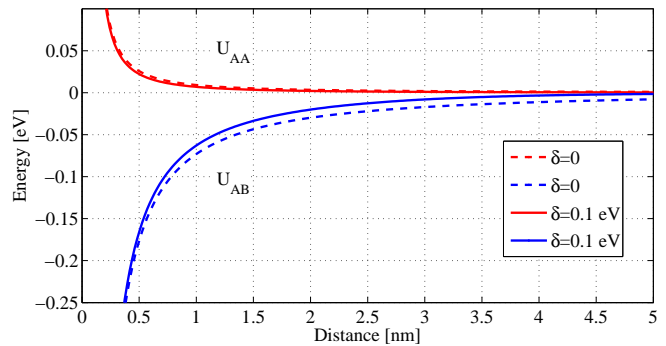}
\caption{The interaction (\ref{eq:AA_final}), (\ref{eq:interaction_AB}) is sensitive to the resonant form of the T-matrix, Eq.(\ref{eq:Tmatrix}). For a nonzero detuning $\delta$, the interaction retains the $1/r$ form at distances $r\lesssim \hbar v_0/\delta$, decreasing more rapidly at larger $r$.  When the system is doped away from neutrality, similar behavior is expected at distances shorter than the Fermi wavelength, $r\lesssim \lambda_{\rm F}$.}
\label{fig3}
\end{figure}


Next, we analyze interaction in a system of adatoms at a finite concentration.
Since electronic states with wavelengths exceeding the distance between adatoms, $\lambda\gtrsim d=n^{-1/2}$, 
are strongly perturbed by scattering, this interaction is of non-pairwise character. For relatively high densities, $n>\delta^2/\hbar^2 v_0^2$, the interaction can be estimated 
using the results for $\delta=0$. This gives an energy $\epsilon_*=\hbar v_0/d$ per adatom, leading to the $n^{3/2}$ scaling for the energy density {\it vs.} adatom concentration, Eq.(\ref{eq:E(n)}).

This behavior was confirmed by numerical analysis of the tight-binding problem (\ref{eq:H}), whereby adatoms were modeled by a local potential taking values exceeding $t_0$. Given a random configuration of $N$ adatoms, we diagonalize the Hamiltonian and sum all negative eigenvalues to evaluate the total energy, $E(N)=\sum_{\epsilon_\alpha<0}\epsilon_\alpha$. The dependence on $N$ is dominated by a contribution linear in $N$, 
$E(N) = E_0 + A_0 N + A_1 N^{3/2}$,
which represents a chemical potential of an adatom. Subtracting the linear part, we recover the interaction
$\Delta E(N)\propto N^{3/2}$ (see Fig.\,\ref{fig2}). Alternatively, one can choose to evaluate $E(N)$ as a sum over the lower half of the spectrum. This changes somewhat the linear term, leaving the $N^{3/2}$ contribution essentially the same.

The sign of interaction is that of {\it attraction} when adatoms are 
evenly spread over both sublattices.  In this case, the best-fit value of the prefactor in the scaling relation (\ref{eq:E(n)}) is found to be $\epsilon_0\approx 0.42\,t_0$. With $t_0=3.1\,{\rm eV}$ this gives $\epsilon_0\approx 1.3\,{\rm eV}$. 
In contrast, when all adatoms are placed on one sublattice, a {\it repulsive} interaction is found,
$\epsilon_0\approx -0.24\,t_0=-0.75\,{\rm eV}$. This is in agreement with the signs of pairwise interaction discussed above.

To test these numerical results against an analytic approach, we use  disorder-averaged Greens functions found in a selfconsistent mean-field approximation, in which point-like adatoms are replaced by a constant field:
%
\be\label{eq:selfconsistency}
\tilde G^{-1}(i\epsilon,\vec k)=i\tilde\epsilon-\lp\begin{array}{cc}
         0 & t_{\vec k}  \\
         t^*_{\vec k} &  0
      \end{array}\rp
,\quad
i\tilde\epsilon=i\epsilon-\frac{\pi v_0^2 n_1}{i\tilde\epsilon\ln\frac{W}{|\tilde\epsilon|}}
,
\ee
where $n_1=2n/3^{3/2}a^2$ is adatoms' density per sublattice, $a=1.42\,{\rm\AA}$ is carbon spacing. Solving the selfconsistency condition (\ref{eq:selfconsistency}) with logarithmic accuracy, we find
\be
\tilde\epsilon=\frac{\epsilon}2+{\rm sgn}\,\epsilon\sqrt{\frac{\epsilon^2}4+\Delta^2}
,\quad
\Delta^2\ln\frac{W}{\Delta}=\pi v_0^2 n_1
.
\ee
The energy density of the system can be written as
\be
E = \oint\frac{dz}{2\pi i} z\sum_\alpha\frac1{z-\epsilon_\alpha}
=\int_{-\infty}^\infty \frac{d\epsilon}{2\pi}i\epsilon\,{\rm Tr}\,G(i\epsilon)
\ee
where $\epsilon_\alpha$ is the spectrum, and the contour integral is taken over the imaginary axis and a half-circle at infinity. The trace of $G$ is identical to that in the self-energy of a T-matrix, giving ${\rm Tr}\,G(i\epsilon)=-2i\tilde\epsilon\ln(W/|\tilde\epsilon|)/\pi v_0^2 $. Subtracting the contribution due to free Dirac fermions, we obtain the change in total energy due to adatoms,
\be 
E_{\rm int}=
\int_{-\infty}^\infty \frac{d\epsilon}{(\pi v)^2} \epsilon\lp
\tilde\epsilon\ln\frac{W}{|\tilde\epsilon|}
-
\epsilon\ln\frac{W}{|\epsilon|}
\rp
\ee
The function under the integral
is even, positive, and approximately constant at $|\epsilon|\gtrsim \Delta$, taking on a value proportional to $n$ (with logarithmic corrections). At $0<\epsilon\lesssim \Delta$ the function is increasing from zero to the asymptotic value at large $\epsilon$.
This behavior is in agreement with expectation of a leading contribution $\delta E\propto n$ and a {\it negative} $n^{3/2}$ part describing interaction. Subtracting the part linear in $n$, and dividing by the density of carbon atoms $n_0$, we find the interaction energy 
\be\label{eq:Omega_int}
\Delta E_{\rm int}=-\frac{8\Delta^3}{3\pi^2v_0^2n_0}
\lp\ln\frac{W}{\Delta}-\frac23\rp
,\quad 
n_0=\frac4{3^{3/2}a^2}
,
\ee
per carbon atom. This formula agrees very well with our numerical results (see red curve in Fig.\ref{fig2}).

A long-range attraction between adatoms can drive thermodynamic instability. This can be seen most easily from phase diagram, obtained from the free energy 
$F=E(n)-TS(n)$ (see Fig.\ref{fig2} inset). In our case,
\be\label{eq:free_energy}
F =-\epsilon_0 n^{3/2}+T\lp n\ln n+(1-n)\ln(1-n)\rp
,
\ee
giving the critical temperature $T_*=\epsilon_0/2\sqrt{3}\approx 4200\,{\rm K}$. 
Since temperature during hydrogenation is substantially 
below $T_*$ \cite{Elias08}, 
the adatoms are expected to self-organize into high and low-density droplets.

Even if spatial diffusion of hydrogen is slow, as may be the case in  \cite{Elias08}, 
initial stages of self-organization terminated by freezing in a low-temperature state would result in macroscopic inhomogeneities.
Such inhomogeneities of the hydrogenated state were indeed observed in the TEM diffraction images described in Ref.\cite{Elias08}. It was also noted that dehydrogenation restores homogeneity, pointing to an intrinsic character of this effect.

The attraction between ``frozen'' adatoms would create a lateral stress. Treating the occupancy $n$ as strain-independent, we have
\be
\sigma=-\partial E(n)/\partial \ln V \approx \frac12|\epsilon_0| n^{3/2}
,
\ee
where an empirical relation
$\partial t_0/\partial a\approx -t_0/a$ is used to describe the change in $t_0$.
Such stress would lead to compression of the graphene matrix. This is consistent with the reduction in lattice period upon hydrogenation observed in experiment \cite{Elias08}.



We are grateful to A. K. Geim, R. L. Jaffe, and K.~S.~Novoselov for useful discussions.


\begin{references}
\vspace{-7mm}

\bibitem{Geim07}
A. K. Geim, K. S. Novoselov, 
Nat. Mater. {\bf 6}, 183 (2007).

\bibitem{Stankovich06} S. Stankovich {\it et al.}, J. Mater. Chem. {\bf 16}, 155 (2006)

\bibitem{Liu08} L. Liu {\it et al.}, 
Nano Lett. {\bf 8}, 1965 (2008).

\bibitem{Ryu08}
S. Ryu {\it et al.}, 
Nano Lett. {\bf 8}, 4597 (2008).


\bibitem{Elias08} D. C. Elias {\it et al.}, 
Science {\bf 323}, 610 (2009).

\bibitem{Duplock04}
E. J. Duplock, M. Scheffler, and P. J. D. Lindan,
Phys. Rev. Lett. {\bf 92}, 225502 (2004).

\bibitem{Sofo07} J. O. Sofo, A. S. Chaudhari, G. D. Barber, 
Phys. Rev. B {\bf 75}, 153401 (2007).

\bibitem{Boukhvalov08} 
D. W. Boukhvalov, M. I. Katsnelson, A. I. Lichtenstein, 
Phys. Rev. B {\bf 77}, 035427 (2008). 


\bibitem{Pepin01} 
C. P\'epin and P. A. Lee, Phys. Rev. B {\bf 63}, 054502 (2001).


\bibitem{Bena05}
C. Bena and S. Kivelson, Phys. Rev. B {\bf 72}, 125432 (2005).


\bibitem{Pereira06} V. M. Pereira {\it et al.}, 
Phys. Rev. Lett. {\bf 96}, 036801 (2006).


\bibitem{Crommie93} 
M. F. Crommie, C. P. Lutz, and D. M. Eigler, Nature (London) {\bf 363}, 524 (1993).

\bibitem{Avouris94} 
P. Avouris, Solid State Commun. {\bf 92}, 11 (1994).

\bibitem{Repp00} J. Repp {\it et al.}, 
Phys. Rev. Lett. {\bf 85}, 2981 (2000).


\bibitem{Einstein73}
T. L. Einstein and J. R. Schrieffer, Phys. Rev. B {\bf 7}, 3629 (1973).

\bibitem{Lau78}
K. H. Lau and W. Kohn, Surf. Sci. {\bf 75}, 69 (1978).


\bibitem{Cheianov06} V. V. Cheianov and V. I. Falko,
Phys. Rev. Lett. {\bf 97}, 226801 (2006). 


\bibitem{Silly04}
F. Silly {\it et al.}, 
Phys. Rev. Lett. {\bf 92}, 016101 (2004).

\bibitem{Zhabinskaya08}
D. Zhabinskaya, J. M. Kinder, E. J. Mele, Phys. Rev. A{\bf 78}, 060103(R) (2008).






\bibitem{AGD}
A. A. Abrikosov, L. P. Gor'kov and I. E. Dzyaloshinsky, Methods of Quantum Field Theory in Statistical Physics (Dover, New York, 1975).


\bibitem{Emig07} T. Emig, N. Graham, R. L. Jaffe, and M. Kardar, Phys. Rev. Lett. {\bf 99}, 170403 (2007).

\bibitem{Kenneth08} O. Kenneth, I. Klich, Phys. Rev. B {\bf 78}, 014103 (2008).

\bibitem{Hornekaer06} L. Hornekaer {\it et al.}, Phys. Rev. Lett. {\bf 96}, 156104 (2006).






\end{references}
\end{document}